\documentclass[preprint,showpacs]{revtex4}
\usepackage{graphicx}
\begin{document}

\title{Logarithmic coarsening and glassy behavior in a
polymer model with mass-dependent diffusion}
\author{F. D. A. Aar\~ao Reis${}^{1,}$\footnote{Email address:
reis@if.uff.br} and R. B. Stinchcombe${}^{2,}$\footnote{E-mail address:
r.stinchcombe1@physics.ox.ac.uk}}
\affiliation{
${}^{1}$ Instituto de F\'\i sica, Universidade Federal Fluminense,
Avenida Litor\^anea s/n, 24210-340 Niter\'oi RJ, Brazil\\
${}^{2}$ Rudolf Peierls Centre for Theoretical Physics,
Oxford University, 1 Keble Road, Oxford OX1 3NP}
\date{\today}

\begin{abstract}
We present a model of polymer growth and diffusion with frustration
mechanisms for density increase and with diffusion rates of Arrhenius form
with mass-dependent energy barriers $\Gamma (m) \sim (m - 1)^{\gamma}$.
It shows non-universal logarithmic coarsening involving the exponent
$\gamma$. Strong-glass behavior is found in the typical times for
disappearance of all polymers up to a given length, without reference to the
equilibrium states of the macroscopic system. These features are predicted by
numerical simulations, scaling theories and an analytic solution of the master
equation within an independent interval approximation, which also
provides the cluster size distribution.
\end{abstract}


\maketitle
\narrowtext

\section{Introduction}

One of the most remarkable features of glassy systems is the rapid increase
of the relaxation time $\tau$ to equilibrium states as the temperature $T$
is lowered. In the most simple glasses, called strong glasses, an Ahrrenius
behavior $\tau\sim\exp{\left( A/T\right)}$ is observed, while in fragile
glasses more complex temperature-dependences are found. This slow relaxation
is expected to be accompanied by a slow growth of correlated domains.
Several microscopic models have already been proposed to represent such
features~\cite{ritort,cugliandolo}. For instance, strong glass behavior was
found in the spin-facilitated models introduced by Fredrickson and
Andersen~\cite{fa,crisanti} and several glassy features were found in models
with kinetic constraints~\cite{ka,toninelli,garrahan,pan}, in which initial
random systems evolve to equilibrium states via the slow diffusion of
deffects. On the other hand, fragile glass
relaxation to equilibrium states with $\tau\sim\exp{\left( B/T^2\right)}$) was
recently shown in a spin chain with asymmetric kinetic
constraints~\cite{jackle,sollich}. The dynamic rules assumed during the
non-equilibrium evolution of these systems are derived from
statistical equilibrium conditions. For instance, in the
spin-facilitated models, the final concentration of defects is given by
the canonical distribution $\exp{\left( -E/T\right)}$ at low temperatures and
this is the origin of the Arrhenius form of the relaxation time in those
systems.

An alternative scenario for the onset of anomalous coarsening and glassy
behavior is suggested in this paper with the analysis of a polymer growth
model in one dimension. The slow dynamics in this model is a consequence of
the interplay between slow activated (Arrhenius) diffusion of clusters and
frustration of density increase. Cluster diffusion occurs in thermal contact
with the surroundings, with energy barriers increasing with cluster length.
Density increase is represented by the deposition of new particles in the
line, but it is not allowed at small vacancies between the clusters. Thus, 
the dynamic rules of this model do not make any reference to equilibrium
macroscopic states and the physical motivation of those processes
contrasts with the somewhat artificial stochastic rules of other simple models
with similar dramatic slowing.

The rules of the model, illustrated in Figs. 1a and 1b, prescribe as
follows the influence of polymer length on activated diffusion, the
suppression of density increase, and the irreversible polymer aggregation. 
A cluster (polymer) can move one lattice spacing to the right or to the left
with diffusion rate given by $r\equiv exp\left( -\Gamma/T\right)$, where
the energy barrier $\Gamma$ increases with polymer length and $T$ is the
temperature (Fig. 1a). We will assume that $\Gamma\sim
\left( m-1\right)^\gamma$, where $m$ is the polymer length (mass) and $\gamma
>0$. Deposition of hard core particles, which represents density increase, is
allowed only at sites with one empty nearest neighbor, with deposition rate
$F=1$ (Fig. 1b). Aggregation of a particle to a cluster and of two
clusters is irreversible and occurs upon any contact of nearest neighbors.
Thus, after a diffusion event, the most typical situation is the
formation of a larger cluster (Fig. 1a). This represents an ideal
polymerization process, with no energy barrier for the formation of a new bond
between neighboring particles and an infinite barrier for
the reverse process. The linear increase of $\Gamma$ with $m$ ($\gamma=1$)
would correspond to polymers stretched along a surface, with $\Gamma$ being
the sum of adsorption energies of all monomers. Although not directly related
to the present problem, it is also relevant to recall that energy barriers
with $\gamma\approx 0.5$ are found in desorption of linear alkanes from
graphite surfaces~\cite{paserba}. Consequently, our numerical study will be
concentrated in systems with $\gamma\sim 1$. 

\begin{figure}[!h]
\includegraphics[clip,width=0.80\textwidth, 
height=0.40\textheight,angle=0]{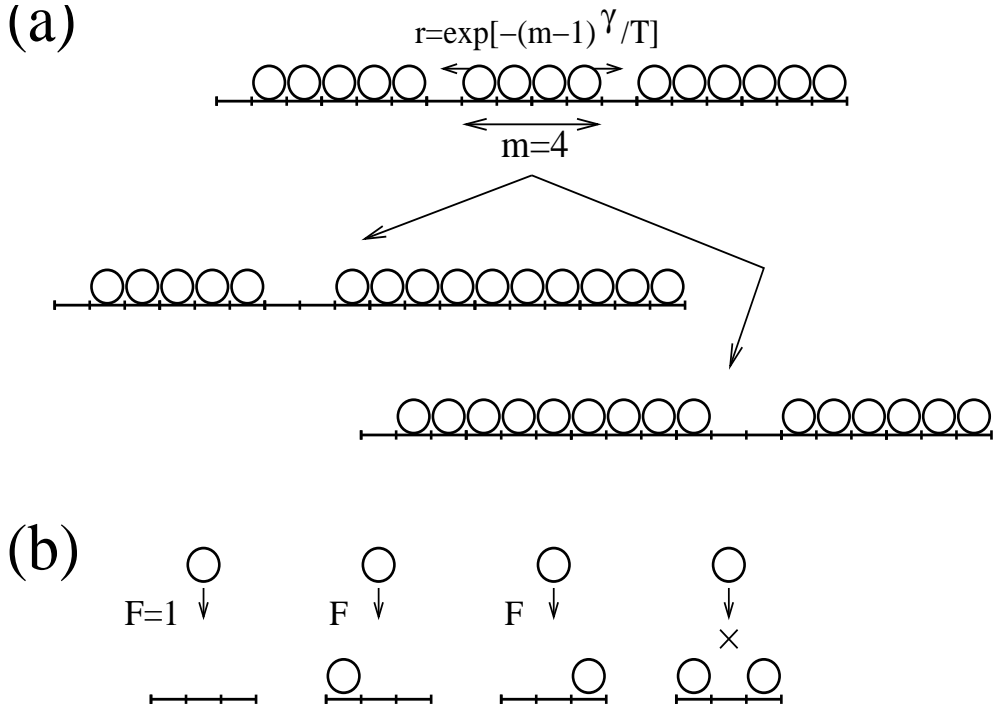}
\caption{\label{fig1} 
(a) Diffusion of a polymer of length (mass) $m$, with the
corresponding rate $r$, and the two possible final configurations after the
polymer move. (b) Allowed deposition processes at vacancies with a
neighboring vacant site, with the corresponding rate, and the forbidden
deposition process, in which the vacancy has two occupied neighbors.}
\end{figure}

The restriction of the model to a low-dimensional structure helps to obtain
the analytical and numerical solutions. This is certainly a simplification
for many systems, in which the dimensionality may play an important
role. However, glassy behavior has been seen in several low-dimensional
systems, such as polymer films~\cite{baljon}, and the formation of clusters of
fast-moving particles with dimensions near or below $2$ was observed in
colloidal glasses~\cite{weeks}. Consequently, we believe that the qualitative
scenario introduced here may find applications to real systems.

Using scaling arguments and numerical methods, we will show that the model
presents non-universal logarithmic coarsening, in which the average cluster
(polymer) length slowly increases as $\langle m\rangle \sim {\left(
T\ln{t}\right)}^{1/\gamma}$, starting from an initial random low-density
configuration. The model also presents strong-glass features of the
characteristic times for eliminating structures of a given length: after a time
$\tau\sim\exp{\left[ d^\gamma/T\right]}$, all clusters of sizes of order $d$
or smaller will disappear. Consequently, the anomalous coarsening and the
Arrhenius-type relaxation are observed during the non-equilibrium system
evolution, which involves frustration of density increase and thermally
activated diffusion at the microscopic level. We will also solve the
master equation of the process within an independent interval approximation,
using the same mapping onto a column picture adopted in the study of related
systems in Refs. \protect\cite{coarsen1} and \protect\cite{coarsen2}. However,
the treatment of cluster size distributions of the present model has novel
and less trivial aspects. It will confirm the above results and provide cluster
length distributions that remarkably differ from those related systems
and that, as far as we know, were not previously obtained in other systems
with logarithmic coarsening. 

The rest of this paper is organized as follows. In Sec. II we use scaling
arguments to predict the logarithmic coarsening and the glassy behavior and
confirm these results with numerical simulations. In Sec. III we solve the
master equation within an independent interval approximation, providing the
cluster length distributions. In Sec. IV we summarize our results and
conclusions.

\section{Scaling of cluster length and glassy behavior}

Scaling arguments can be used to predict the slow coarsening of this model,
following the same lines of Ref. \protect\cite{mevansreview}, which were
previously applied to magnetic systems~\cite{lai,shore} and to related
non-equilibrium models~\cite{coarsen2,coarsen1}. For simplicity, we refer to
the average cluster mass as $m$.

\begin{figure}[!h]
\includegraphics[clip,width=0.70\textwidth, 
height=0.40\textheight,angle=0]{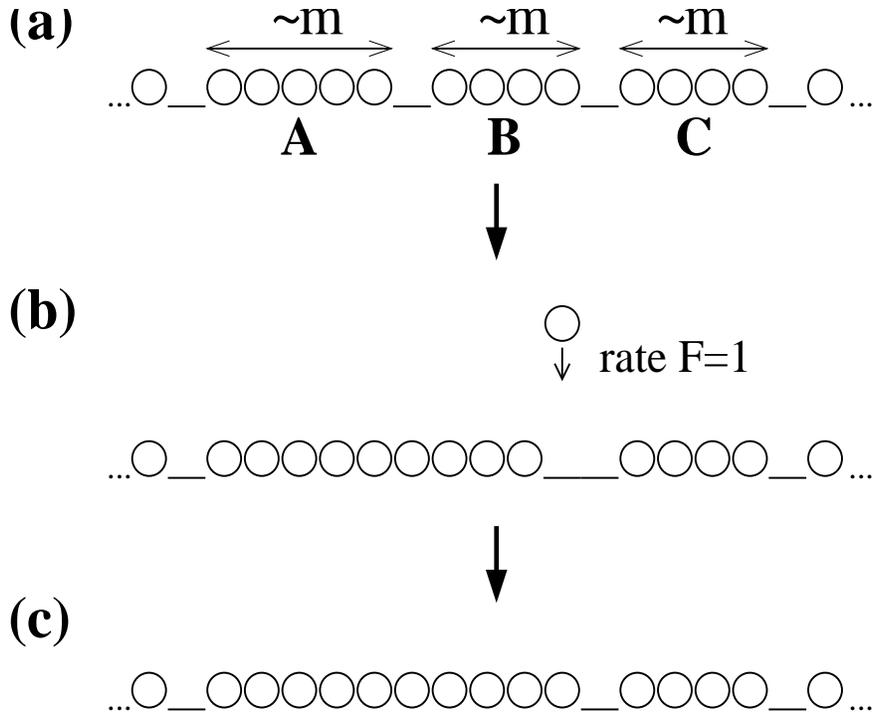}
\caption{\label{fig2} 
Illustration of the coarsening dynamics for the construction of the
scaling theory: (a) three typical neighboring clusters are shown, with lengths
of the order of the average cluster size $m$; (b) cluster $B$ moves to
the left, forming a larger cluster after aggregation to cluster $A$, and a
new particle is deposited at one of the sites of the double vacancy between
$A+B$ and $C$; (c) the final configuration after the deposition of a new
particle at the left site of the double vacancy.}
\end{figure}

In Fig. 2a, we show a configuration with
clusters of lengths typically of order $m$, named $A$, $B$ and $C$, with
single empty sites between them. The time necessary for cluster $B$ to move is
of order $\Delta t = \exp{\left(m^\gamma/T\right)}$. If it moves to the left
(Fig. 2b), then clusters $A$ and $B$ coalesce. Since the diffusion rates of
clusters $A+B$ and $C$ are very small for large $m$, while the deposition rate
is $F=1$, a new particle will immediately be deposited in one of the sites of
the double vacancy (Fig. 2b). One possible configuration after the deposition
of a new particle is shown in Fig. 2c. From the initial to the final
configuration (Fig. 2a to Fig. 2c), the average cluster length increase was of
order $m$. Thus we obtain
\begin{equation}
{{dm}\over{dt}} \sim {{\Delta m}\over{\Delta t}} \sim
{m\over{\exp{\left(m^\gamma/T\right)}}} .
\label{scalingeq}
\end{equation}

We will be mainly interested in the long-time regime in which cluster
diffusion is sufficiently slow, even for small $\gamma$. In this case, the
exponential factor at the right-hand side of Eq. (\ref{scalingeq}) is much
larger than $m$, which may be neglected at first approximation. Thus,
integrating Eq. (\ref{scalingeq}), we are led to the scaling of the average
cluster length as
\begin{equation}
m \sim {\left( T\ln{t}\right)}^{1/\gamma} .
\label{logcoarsen}
\end{equation}
The leading correction due to the neglected term in Eq. (\ref{scalingeq}) is
proportional to $\ln{\left(
\ln{t}\right)}/{\left(\ln{t}\right)}^{1-1/\gamma}$. However, it is not expected
to be the true leading correction to the dominant scaling of the model because
Eq. (\ref{scalingeq}) is itself an approximation, which omitted further
corrections.

Numerical simulations of this model confirm the logarithmic coarsening.
In Fig. 3 we show the time evolution of
the average cluster length for three sets of values of $\gamma$ and $T$, in
which that scaling is observed in up to $7$ decades of time. The data for
$\gamma=1$, with $T=1$ and $T=2$, are averages over $3\times {10}^4$ different
realizations in lattices of length $L={10}^4$.
The data for $\gamma=0.5$, in which much larger
clusters appear at small times, are averages over $200$ realizations in
lattices with $L=5\times {10}^4$, up to time $t={10}^5$.

\begin{figure}[!h]
\includegraphics[clip,width=0.69\textwidth, 
height=0.46\textheight,angle=0]{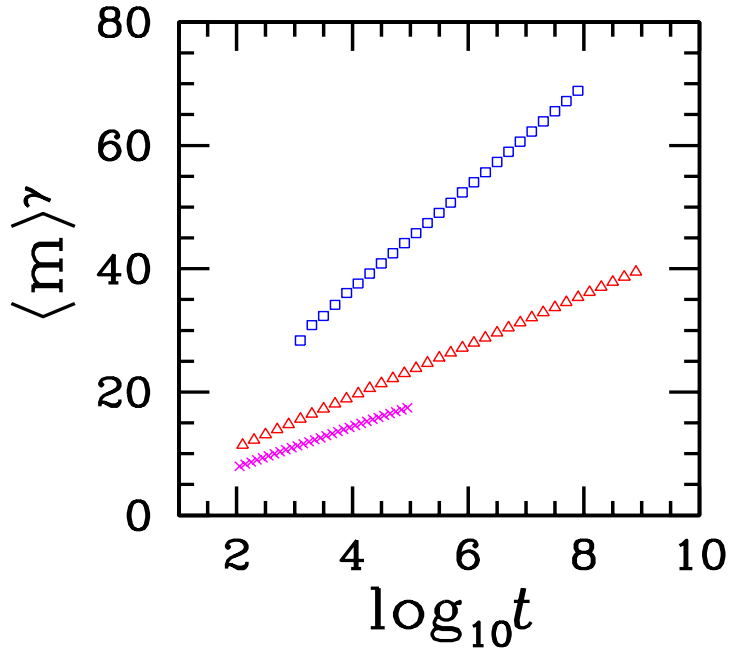}
\caption{\label{fig3}  ${\langle m\rangle}^\gamma$ versus
$\log{\left( t\right)}$ for: $\gamma=1$ and $T=1$ (squares);
$\gamma=1$ and $T=2$ (triangles); $\gamma=0.5$ and $T=1$ (crosses).}
\end{figure}

In Fig. 4 we show the data for the same values of $T$ and $\gamma$ scaled in
the form $\langle m\rangle /{\left( T\ln{t}\right)}^{1/\gamma}$ versus
$1/\ln{t}$. As $t\to\infty$ ($1/\ln{t}\to 0$), Fig. 4 suggests an universal
amplitude in the scaling relation (\ref{logcoarsen}), although corrections to
the dominant scaling clearly depend on $\gamma$ and $T$.

\begin{figure}[!h]
\includegraphics[clip,width=0.69\textwidth, 
height=0.46\textheight,angle=0]{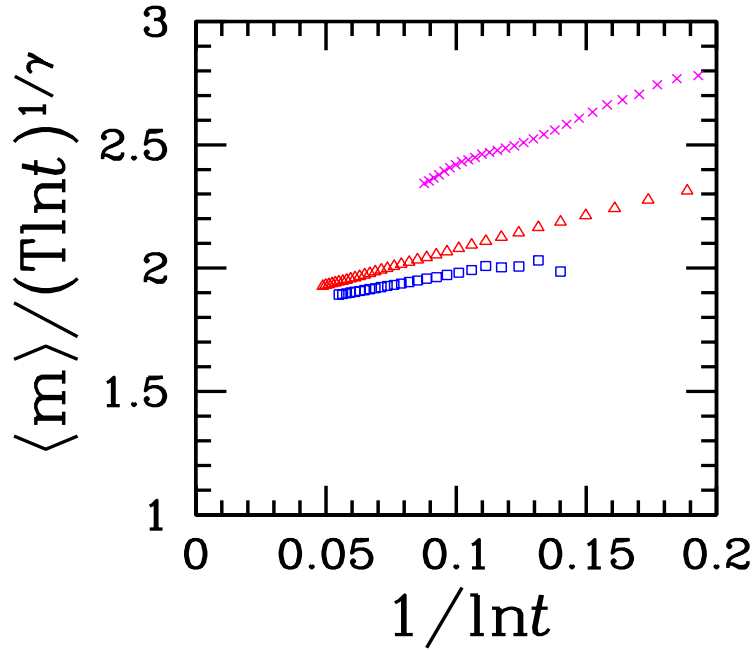}
\caption{\label{fig4}
Scaled average cluster size as a function
of inverse of $\ln{t}$ for: $\gamma=1$ and $T=1$ (squares);
$\gamma=1$ and $T=2$ (triangles); $\gamma=0.5$ and $T=1$ (crosses).}
\end{figure}

The logarithmic coarsening leads to a time evolution of the density with a
logarithmic correction, $\rho =1 - 1/{\left(
T\ln{t}\right)}^{1/\gamma}$. This type of time-dependence was also found in
experiments of compaction of granular systems~\cite{knight} and theoretical
models to describe them~\cite{robinprl,lefevre}, but temperature played no
role there.

Another remarkable feature of this system seen in simulations is the
absence of clusters whose reduced lengths
\begin{equation}
y\equiv m/{\left( T\ln{t}\right)}^{1/\gamma}
\label{defy}
\end{equation}
are smaller than
the most probable value $y=a$ ($a\to const$ for large $t$). In order
to understand this feature, consider a set of clusters with reduced size
$y=a-\epsilon$, with any $\epsilon>0$, whose diffusion rate is
$t^{{\left( a-\epsilon\right)}^\gamma}$. This value is larger than the
diffusion rate of the clusters with $y=a$ by a factor $t^{\epsilon\gamma/a}$.
Thus, at large $t$, compared to the most probable clusters, the clusters with
$y=a-\epsilon$ are much more mobile and are present in smaller number in the
line. So they will merge into the slower larger neighboring clusters almost
instantaneously within the typical time scale for diffusion of the most
probable clusters (length $y=a$), which causes the removal of clusters with
$y=a-\epsilon$.

This is consistent with the simulation results shown in Fig. 5,
in which we plotted the reduced probability of cluster
length $m$, ${\left( T\ln{t}\right)}^{1/\gamma} P\left( m\right)$, as a
function of the reduced length $y$, for $\gamma =1$ (two temperatures in the
main plot) and $\gamma=0.5$ (in the inset).
Below a certain value $y=a\approx 1$, the corresponding
probabilities rapidly decrease to zero and the cluster size distribution is
asymptotically discontinuous at that point. A discontinuity in the slope of
the distribution at $y\approx 2a$ is also suggested in Fig. 5 and will be
rigorously justified in Sec. III. Further slope discontinuities will be
predicted, but they are not so clear in Fig. 5 due to the larger fluctuations
in the data for large $y$.

\begin{figure}[!h]
\includegraphics[clip,width=0.69\textwidth, 
height=0.46\textheight,angle=0]{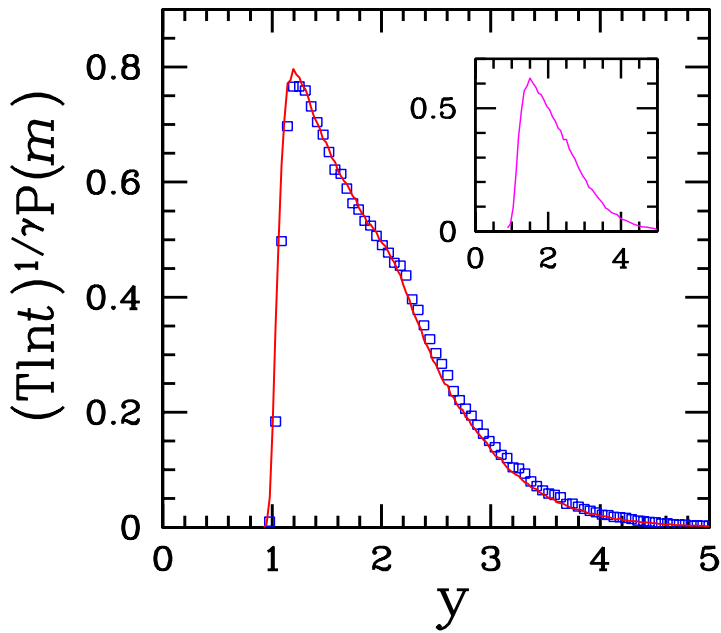}
\caption{\label{fig5}
Scaled probability of clusters of length
(mass) $m$ as a function of the scaled length $y$ for $\gamma=1$, with $T=1$
(solid curve) and $T=2$ (squares), at $t={10}^8$. The inset shows the same
quantities for $\gamma =0.5$ and $T=1$, at $t={10}^7$ (not shown in the main
plot to avoid superposition of many data points).}
\end{figure}

From the above results, the characteristic time for clusters of a given mass
$m$ to be completely eliminated from the system is
$t=\exp{\left[ {\left( m/a\right)}^\gamma /T \right]}$. 
The connection of the glassy behaviour to the thermally activated diffusion
of microscopic structures is clear, in contrast to other kinetically
constrained models whose dynamic rules are determined by the statistical
equilibrium of the system.

\section{Analytical solution}

Now we turn to the analysis of the model starting from a
version of the master equation, following the same lines as in the solution of
related models in Refs. \protect\cite{coarsen1} and
\protect\cite{coarsen2}.

In our original problem, each site on the line is occupied by one or zero
particles, but we are interested in the evolution of cluster length. Thus, the
analysis of a master equation is more easily set up by reformulating the
process using a column picture, in which a column of height $m$ represents a
cluster of size $m$ together with its adjacent vacancy on the right.
This mapping is illustrated in Fig. 6.

\begin{figure}[!h]
\includegraphics[clip,width=0.70\textwidth, 
height=0.35\textheight,angle=0]{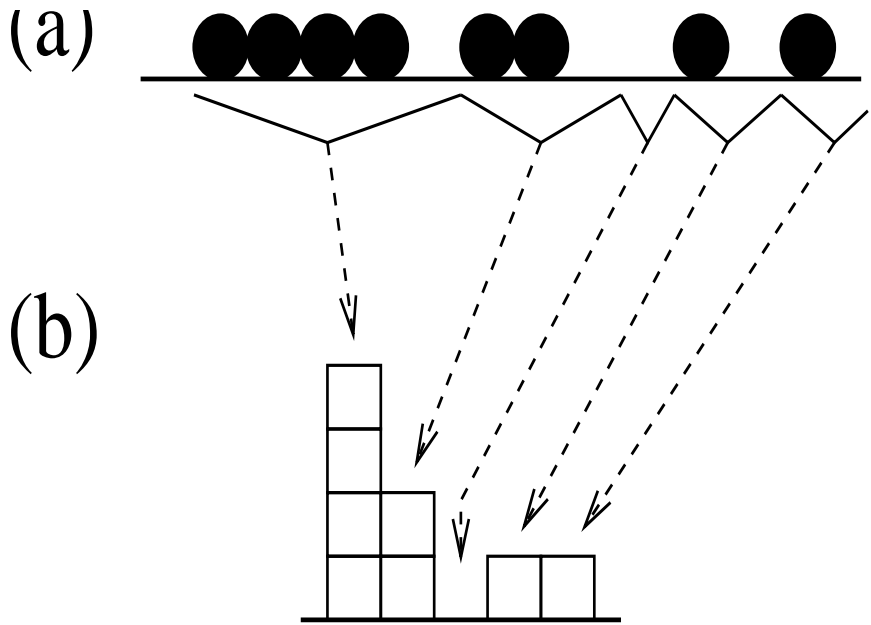}
\caption{\label{fig6}
Example of particle-hole configuration on a line (a) and the map
(dashed arrows) into a column problem (b). Each cluster and the vacancy at its
right side in (a) correspond to a column in (b) with the same mass.}
\end{figure}

Fig. 6 clearly shows that the lattice
length $L_0$ in the column picture is smaller than the length $L$ in the
original problem (particle picture): $L_0=L-M$, where $M$ is the total mass,
for periodic boundaries. As time increases and new particles are deposited,
$L_0$ decreases. However, new particles are deposited only at double vacancies
of the particle picture (Fig. 1b) and the latter correspond to single vacancies
in the column picture (Fig. 6). Thus, the decrease of $L_0$ (or mass increase)
is related to the probability $P_t(0)$ of an empty site in the column problem
as ${{L_0(t+1)-L_0(t)}\over{L_0(t)}} = -P_t(0) \left[ 2-P_t(0)\right]$.

The probability $P_t(m)$ that a randomly chosen cluster
in the column picture has size $m$ at time $t$ is given in terms of
the clusters numbers, $N(m,t)$, by $P_t(m) = {{N(m,t)}\over{L_0(t)}}$.
The gain/loss from in/out processes provides a master equation for
clusters numbers as
\begin{equation}
N(m,t+1) - N(m,t) = L_0 \left( {\cal B}_m + E_m \right) ,
\label{masternumbers}
\end{equation}
where ${\cal B}_m$ comes from the deposition processes and $E_m$ comes from
diffusion of clusters or isolated particles. It can be rewritten in terms of
cluster length probabilities as
\begin{equation}
{\left( 1-P_t(0)\right)}^2 P_{t+1}\left( m\right) - P_t\left( m\right) =
{\cal B}_m + E_m ,
\label{masterequation}
\end{equation}
which allows for the changes in $L_0$.

In an independent interval approximation (IIA) in which joint probabilities are
factorized, the contribution from deposition is given by
\begin{equation}
{\cal B}_m = P_t(0) \left[ 2 \theta\left( m-2\right) P_t\left( m-1\right) -
2 \theta\left( m-1\right) P_t\left( m\right) + \delta_{m,1}P_t\left( 0\right)
- 2\delta_{m,0} \right] .
\label{masterdep}
\end{equation}
Here, the factor $P_t(0)$ accounts for the fact that deposition is allowed
only at single vacancies in the column picture (Fig. 6).  The first term
inside brackets gives the rate of production of clusters of size $m$ from
deposition at one of the sides of a cluster of size $m-1$, for $m\geq 2$,
while the third term gives the rate of production of a cluster with
$m=1$ in a double vacancy (triple vacancy in the particle picture, with the new
particle being deposited at the central site). The second term corresponds to
the loss of a cluster of mass $m$ due to deposition at one of its sides and
the last term the loss of vacancies due to deposition at one of the two
corresponding sites in the particle picture.

The contribution from diffusion in Eq. (\ref{masterequation}) is given by
\begin{eqnarray}
E_m &=& \sum_{r=1}^{m}{ f(r) P_t(r) P_t(m-r)} +
\delta_{m,0} \sum_{r=1}^{\infty}{ f(r) P_t(r)}
\nonumber\\
&&- \theta (m-1) f(m) P_t(m)
- P_t(m) \sum_{r=1}^{\infty}{ f(r) P_t(r)} .
\label{masterdifpol}
\end{eqnarray}
In Eq. (\ref{masterdifpol}), the function $f(m)\equiv \exp{\left[
-{\left( m-1\right)}^\gamma/T \right]}$ gives
the diffusion rate for a cluster of mass $m$.
The first term in Eq. (\ref{masterdifpol}) gives the rate of production of a
cluster of mass $m$ due to the coalescence of smaller clusters after diffusion
of one of them. The second term corresponds to the creation of a single vacancy
(double vacancy in the particle picture) due to cluster diffusion - see Fig.
1a. The loss terms correspond to coalescence of a cluster of mass $m$ and
a neighboring cluster: the third one accounts for diffusion of the cluster of
mass $m$ and the fourth one accounts for diffusion of the neighboring clusters.

In the scaling limit of large $t$ and large $m$ (coarsening), the probability
$P_t(m)$ must have the form
\begin{equation}
P_t(m) = {1\over{L\left( t\right)}} g\left( y\right) \qquad ,\qquad
y\equiv {m\over {L\left( t\right)}} ,
\label{scalingprob}
\end{equation}
with some function $L(t)$ characterizing the growing typical cluster size.

Figs. 2a-c show that the survival time of a double vacancy is of order
$1/F=1$, while a typical cluster move occurs on a much larger timescale, of
order $\exp{\left( m^\gamma/T \right)}$. Consequently, the onset of a double
vacancy (in the particle picture) is a rare event. Nevertheless, since the
double vacancy mediates deposition, its
probability is required in the equation for the one-variable scaling function
$g$, and the same scaling arguments of Sec. II lead to the expected scaling of
the double vacancy probability in the particle picture as $P_t\left(
0\right) = \left( C/2\right) /t$ (see e.g. Ref. \protect\cite{coarsen2}).

Thus we obtain an equation for the scaling function as
\begin{eqnarray}
-\frac{L'}{L} \left[ g + y \frac{dg}{dy} \right] - \frac{Cg}{t}
&=&
- \frac{C \delta\left( y\right)}{t} 
-h(y) - g(y)I(\infty ) +
\delta\left( y\right) I(\infty ) +
\nonumber\\
&& \int_0^y{dy' g\left( y-y'\right) h\left( y'\right) } ,
\label{eqg}
\end{eqnarray}
where $h(y) = g(y) \exp{\left[ -{\left( Ly\right)}^\gamma/T \right]}$ and
$I(\infty ) = \int_0^\infty{dy' g\left( y'\right)
\exp{\left[ -{\left( Ly'\right)}^\gamma /T\right]} }$.
The first term on the left-hand side (LHS) of Eq. (\ref{eqg}) follows from
those in Eq. (\ref{masterequation}) in the long time, continuous limit. The
second term on the LHS and the first one on the right-hand side
(RHS) correspond to deposition (Eq. \ref{masterdep}) and the other terms on the
RHS correspond to cluster diffusion (Eq. \ref{masterdifpol}).

Eq. (\ref{eqg}) and the correponding equation for the generating function
contrast to those of the models of Refs. \protect\cite{coarsen1} and
\protect\cite{coarsen2} because a simple power-counting in Eq. (\ref{eqg})
is not enough to provide the scaling of the average cluster size here.
From the structure of Eq. (\ref{eqg}) and the forms of $h(y)$ and $I(\infty)$,
we expect that the balance of dominant terms will be possible if
\begin{equation}
\frac{L'}{L} =
\exp{\left[ -{\left[ L\left( t\right) a\right]}^\gamma/T\right]}
\alpha{\left( L\left( t\right)\right)} ,
\label{eql}
\end{equation}
where $a$ is some constant (see below) and the function $\alpha$ accounts for
possible power-law subdominant factors. Notice also that the dominant factors
of Eq. (\ref{eql}) are the same as those of Eq. (\ref{scalingeq}), obtained
from simple scaling arguments.

As $t\to\infty$, we expect $L(t)\to\infty$, but the last term in the
RHS of Eq. (\ref{eqg}) will provide a divergent contribution
for $y'<a$. This contribution is not balanced out by other terms unless
$g(y)=0$ for $y<a$.
With this assumption, we obtain a dominant term on the LHS
of Eq. (\ref{eqg}) with a factor $e^{-\left( La\right)^\gamma/T}$ and,
in the integral at the RHS, a factor $e^{-\left(
Ly'\right)^\gamma/T}$, which provides a dominant cancelling term for $y\geq a$.

The solution of Eq. (\ref{eql}) is exactly
the logarithmic growth of $L(t)$ obtained from scaling arguments (Eq.
\ref{logcoarsen}), thus providing the expected scaling of average cluster
length. The above assumption for the function $g$ confirms, within the
IIA, the elimination of all clusters with masses smaller than $aL(t)\sim
{\left( T\ln{t}\right)}^{1/\gamma}$.

The scaling function has the form
\begin{equation}
g(y) = \theta\left( y-a\right) g_1(y) ,
\label{defg1}
\end{equation}
which leads to an equation for the
function $g_1(y)$ as
\begin{equation}
g_1 + y\frac{dg_1}{dy} + Cg_1 -C\delta (y) = \frac{ag_1\left( a\right)}{k}
\left[ g_1\theta (y-a) - g_1(y-a)\theta (y-2a) + \delta (y-a) -
\delta (y) \right] ,
\label{eqg1}
\end{equation}
where $k$ is constant. Consequently, a discontinuity in the slope of $g(y)$ is
obtained at $y=2a$. Subsequently, further discontinuities appear at
$y=3a$, $y=4a$ and so on. The function $g$ can be shown to be continuous
except at $y=a$, where the discontinuity implies $k=1$ in Eq. (\ref{eqg1}).
These findings agree with those in Fig. 5, in which the probabilities of
cluster lengths with $y<a\sim 1$ tend to zero asymptotically and the slope of
the scaled probability clearly changes at $y\approx 2a$.

From Eq. (\ref{eqg1}) we obtain the distribution for $a<x<2a$ as
\begin{equation}
g(y) = A y^{\left[ ag(a)-(C+1)\right]},
\label{decayg}
\end{equation}
where $A$ is a constant.
In Fig. 7 we show $\log{P(m)}$ versus $\log{y}$ for
$\gamma=1$ and $T=1$, at time $t={10}^8$, with a linear fit that
confirms the power-law decay predicted in Eq. (\ref{decayg}). Notice
that, although the scaling arguments of Sec. II were capable of predicting
several features of the model, the analytical solution of the master equation,
even within the IIA, is essential to predict the shape of the scaled cluster
size distribution.

\begin{figure}[!h]
\includegraphics[clip,width=0.69\textwidth, 
height=0.46\textheight,angle=0]{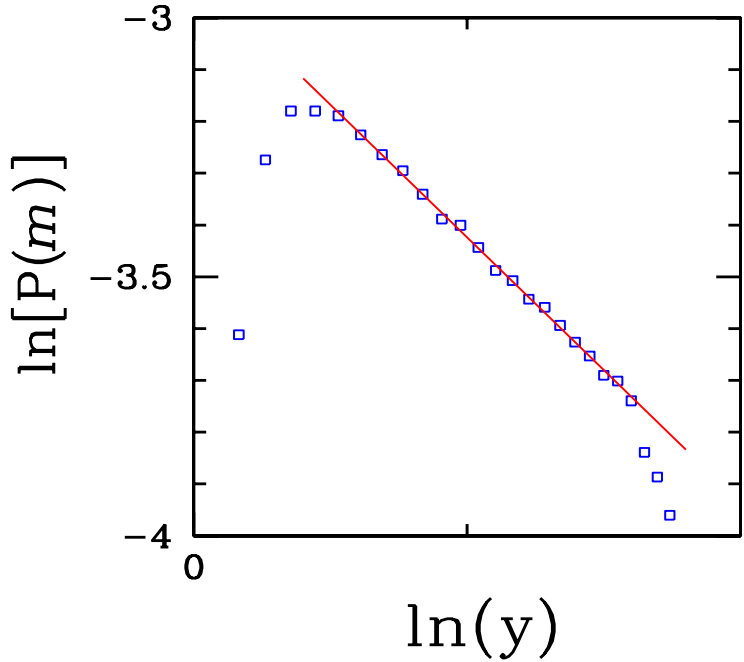}
\caption{\label{fig7}
$\ln{\left[ P\left( m\right)\right]}$ versus
$\ln{\left( y\right)}$ for $\gamma=1$ and $T=1$, at $t={10}^8$, and a linear
fit of the data in the region $1.4\leq y\leq 2.2$.}
\end{figure}

For $2a<y<3a$, the distribution
is $g(y) = B y^{\left[ ag(a)-(C+1)\right]} - A {\left( y-a\right)}^{\left[
ag(a)-(C+1)\right]}$, but the low accuracy of our data in that region does not
allow a reliable test of this form.

\section{Conclusion}

We presented a one-dimensional model of polymer growth and diffusion with
frustration mechanisms for density increase and with diffusion rates of
Arrhenius form, with mass-dependent energy barriers. The non-universal
logarithmic coarsening involves the exponent $\gamma$ and can be predicted by
scaling arguments. The model also shows strong glass behavior of the
characteristic times for elimination of all polymers up to a given length.
These features are confirmed by the solution of the master equation within an
independent interval approximation (IIA). This solution also provides the
distribution of cluster sizes, which shows discontinuities at a sequence of
points and a power-law decay for sizes near the most probable one. These
findings are confirmed by numerical simulations with good accuracy.

The logarithmic coarsening and Arrhenius-type behavior were also obtained
in other systems, but our model presented a different scenario for the
onset of these properties, in which there is no reference to equilibrium
states. Instead, all these features are obtained
during the non-equilibrium system evolution assuming a thermal contact
between the clusters and their surroundings at a temperature $T$, but not an
approach to an equilibrium macroscopic state with that temperature.

The elimination of clusters with lengths below a certain threshold is also
present in the paste-all model of Derrida et al~\cite{derrida}, although
that is a direct consequence of the model prescription, namely the
elimination of the smallest cluster of the system at any time. Instead, in the
present model that elimination is obtained only in the scaling limit (long
times, large clusters), in which diffusion coefficients of small clusters
become infinitely larger than those of large clusters.
The same type of distribution appears in the East model~\cite{jackle,sollich},
in which the rules for spin flips are asymmetric (an artificial but
essential ingredient for obtaining slow dynamics there). On the
other hand, a power-law coarsening and fragile glass behavior
are obtained in the East model, which contrasts to our findings.

The successful application of the IIA for the present problem and for
previous models with particle detachment from
clusters~\cite{coarsen1,coarsen2} also deserves some comments. In these
models, the coarsening arises from coalescence of clusters and the reverse
process is increasingly improbable (altogether improbable in the present
case). Since the initial state is without cluster-cluster correlations, no
correlations between the masses/lengths of neighbouring clusters can build up,
so the IIA becomes exact in the late coarsening limit. This interpretation is
corroborated by the successful comparison of IIA predictions and simulation
results. It seems that IIA fails only when we focus on rare processes which
occur in narrow time windows in which the system is dominated by reversible and
highly correlated processes~\cite{coarsen2}, which is not the case in the
present model.

\begin{acknowledgments}

FDAA Reis thanks the Rudolf Peierls Centre for Theoretical Physics of Oxford
University, where part of this work was done, for the hospitality, and
acknowledges support by CNPq and FAPERJ (Brazilian agencies).

RB Stinchcombe acknowledges support
from the EPSRC under the Oxford Condensed Matter Theory Grants,
numbers GR/R83712/01 and GR/M04426.

\end{acknowledgments}

\end{document}